\documentclass[twocolumn,showpacs,amsmath,amssymb,floatfix,superscriptaddress,letter]{revtex4}

\usepackage{graphicx}
\begin{document}

\hyphenation{}
\def\W{{W^{ij\sigma}_{kl\sigma'}}}
\def\i{{ijkl\sigma\sigma'}}

\title{Performance analysis of continuous-time solvers for quantum impurity models}
\author{Emanuel Gull}
\affiliation{Institut f\"ur theoretische Physik, ETH Z\"urich, CH-8093 Z\"urich, Switzerland}
\author{Philipp Werner}
\affiliation{Columbia University, 538 West, 120th Street, New York, NY 10027, USA}
\author{Andrew Millis}
\affiliation{Columbia University, 538 West, 120th Street, New York, NY 10027, USA}
\author{Matthias Troyer}
\affiliation{Institut f\"ur theoretische Physik, ETH Z\"urich, CH-8093 Z\"urich, Switzerland}

\begin{abstract}
Impurity solvers play an essential role in the numerical investigation of strongly correlated electrons systems within the ``dynamical mean field" approximation. Recently, a new class of continuous-time solvers has been developed, based on a diagrammatic expansion of the partition function in either the interactions or the impurity-bath hybridization. We investigate the performance of these two complimentary approaches and compare them to the well-established Hirsch-Fye method. The results show that the continuous-time methods, and in particular the version which expands in the hybridization, provide substantial gains in computational efficiency. 
\end{abstract}


\maketitle

\section{Introduction}

The numerical investigation of strongly correlated fermion systems on a lattice is a fundamental goal of modern condensed matter physics. Unfortunately, the standard Monte Carlo simulation technique is hampered by the sign problem \cite{Loh90, Troyer05}, which leads to an exponentially growing statistical error with increasing system size or decreasing temperature. Exact diagonalization, on the other hand, is only possible for a small number of sites because of the exponential growth of the Hilbert space \cite{Dagotto91}. 

A useful and numerically feasible approach to treat fermionic systems in the thermodynamic limit is the so-called dynamical mean field theory. Development of this field started with the demonstration by M\"uller-Hartmann and by Metzner and Vollhardt \cite{Mueller-Hartmann89, Metzner89} that the diagrammatics of lattice models of interacting fermions simplifies dramatically in an appropriately chosen infinite dimensional (or infinite coordination) limit. This insight was developed by Georges, Kotliar and co-workers \cite{Georges92, Georges96} who showed that if the momentum dependence of the electronic self-energy may be neglected ($\Sigma(p,\omega)\rightarrow \Sigma(\omega)$), as occurs in the infinite coordination number limit, then the solution of the lattice model may be obtained from the solution of a quantum impurity model plus a self-consistency condition. 

Quantum impurity models are amenable to numerical study. A well-established method is the Hirsch-Fye algorithm \cite{Hirsch86}, which employs a discrete Hubbard-Stratonovich transformation to decouple four-fermion terms. The method however requires the discretization of imaginary time into equal slices, which restricts it to relatively high temperatures. Further, analysis of the ``Slater-Kanamori'' interactions relevant to partially filled d-orbitals requires a very large number of Hubbard-Stratonovich fields \cite{Sakai}. This further complicates the simulations.

Recently, a new class of continuous-time impurity solvers has been developed \cite{Rubtsov05, Werner06, Werner06b}. These diagrammatic QMC approaches rely on an expansion of the partition function into diagrams and the resummation of diagrams into determinants. A local update Monte Carlo procedure is then used to sample these determinants stochastically. Two complementary approaches have been formulated: the weak coupling method \cite{Rubtsov05} uses a perturbation expansion in the interaction part, while the hybridization expansion method \cite{Werner06, Werner06b} treats the local interactions exactly and expands in the impurity-bath hybridization. 
In the weak-coupling case, the determinantal formulation, which eliminates or at least greatly alleviates the sign problem, originates from Wick's theorem. In the hybridization expansion, when starting from a Hamiltonian formulation, the determinants emerge naturally from the trace over the bath states \cite{Werner06b}.  

Both continuous-time methods appear to provide considerable improvements over Hirsch-Fye. 
Our purpose here is to compare the performance of the two continuous-time solvers to each other and to Hirsch-Fye in an objective way. We use the algorithms and measurement procedures as proposed in Refs.~\cite{Hirsch86, Rubtsov05, Werner06}, and focus on the accuracy with which physical quantities can be obtained in a DMFT calculation, for fixed CPU time in the impurity solver step.  

The rest of this paper is organized as follows: section II presents the needed formalism, section III describes aspects of the measurement procedure, section IV gives results and section V is a conclusion.

\section{Theory}
\subsection{Model}
We concentrate in this work on the Hubbard model. For one band, the corresponding single-site impurity model is specified by the imaginary time effective action
\begin{eqnarray}
S_\text{eff} &=& -\int_0^\beta d\tau d\tau' \sum_\sigma c_\sigma(\tau) F_\sigma(\tau-\tau')c_\sigma^\dagger(\tau')\nonumber\\
&& -\int_0^\beta d\tau \Big[\mu(n_\uparrow +n_\downarrow)-Un_\uparrow n_\downarrow\Big], 
\label{S}
\end{eqnarray}
where $\mu$ denotes the chemical potential and $U$ the on-site repulsion. The hybridization function $F$ describes transitions into the bath and back and is related to the mean field function ${\cal G}_0$ by \cite{Werner06b, Georges96}
\begin{equation}
{\cal G}_{0,\sigma}^{-1}(i\omega)=i\omega+\mu-F_\sigma(-i\omega).
\label{G_0} 
\end{equation}
The task of the impurity solver is to compute the Green function
\begin{equation}
G(\tau-\tau')=-\langle T_\tau c(\tau)c^\dagger(\tau')\rangle_{S_\text{eff}}=-\frac{Tr T_\tau e^{-S_{\text{eff}}}c(\tau)c^\dagger(\tau')}{Tr T_\tau e^{-S_{\text{eff}}}}
\label{G}
\end{equation}
for a given hybridization function. 

\subsection{Hirsch-Fye impurity solver}

The algorithm of Hirsch and Fye \cite{Hirsch86} requires a discretization of imaginary time into $N$ slices $\Delta\tau=\beta/N$. In each time slice, the four-fermion term $Un_\uparrow n_\downarrow$ is decoupled using a discrete Hubbard-Stratonovich transformation,
\begin{equation}
e^{-\Delta\tau U(n_\uparrow n_\downarrow+1/2(n_\uparrow+n_\downarrow))}=\frac{1}{2}\sum_{s=\pm 1}e^{\lambda s(n_\uparrow+n_\downarrow)}, 
\label{discreteHS}
\end{equation}
where the parameter $\lambda$ is defined as $\lambda=\cosh(e^{\Delta\tau U/2})$. The Gaussian integral over the fermion fields may then be performed analytically, yielding an expression for the partition function of the form 
\begin{equation}
Z=\sum_{\{s_i\}} \det \Big[ D_{{\cal G}_0^{-1},\uparrow}(s_1,...,s_N)D_{{\cal G}_0^{-1},\downarrow}(s_1,...,s_N) \Big].
\label{Z_HF}
\end{equation}
Here, $D_{{\cal G}_0^{-1},\sigma}(s_1, ..., s_N)$ denotes the $N\times N$ matrix of the inverse propagator for a particular configuration of the auxiliary Ising spin variables $s_1, \ldots, s_N$ \cite{Georges96}. The Monte Carlo sampling then proceeds by local updates in these spin configurations. Each successful update requires the calculation of the new determinants in Eq.~(\ref{Z_HF}), at a computational cost of $O(N^2)$. 

The problem with this approach is the rapid (and, for metals, highly non-uniform) time-dependence of the Green functions at low temperature and strong interactions. The initial drop of the Green function is essentially $\sim e^{-U\tau/2}$, from which it follows that a fine grid spacing $N\sim \beta U$ is required for sufficient resolution. In the Hirsch-Fye community, $N=\beta U$ is apparently a common choice, 
although we will see below that this number is too small and leads to significant systematic errors. As noted in Ref.~\cite{Werner06} a resolution of at least $N=5\beta U$ is typically needed to get systematic errors below the statistical errors of a reasonably accurate simulation.

At half filling, the matrices $D_{{\cal G}_0^{-1},\uparrow }$ and $D_{{\cal G}_0^{-1},\downarrow}$ are identical and it then follows immediately from Eq.~(\ref{Z_HF}) that the Hirsch-Fye algorithm under these conditions does not suffer from a sign problem. In fact, a closer analysis reveals that the sign problem is absent for any choice of $\mu$ \cite{Yoo05}.

\subsection{Weak coupling expansion}

\begin{figure}[ht]
\begin{center}
\includegraphics[width=.4\textwidth]{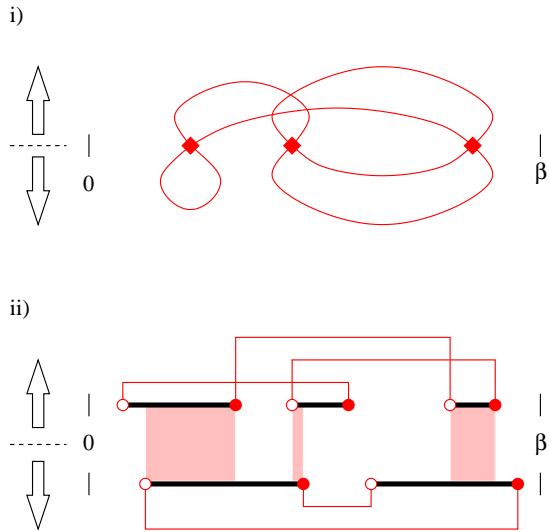} \\ 
\end{center}
\caption[]{Illustration of the diagrams generated by the continuous-time impurity solvers. i) Weak-coupling method: third order diagram consisting of three vertices (diamonds) $Un_\uparrow(\tau)n_\downarrow(\tau)$ linked by lines representing the function ${\cal G}_{0,\sigma}(\tau_i-\tau_j)$. ii) Hybridization expansion method: here, the orders of the diagrams for up- and down-spins can be different. Each creation operator $c^\dagger_\sigma(\tau_s)$ (empty dot) is  connected to an annihilation operator $c_\sigma(\tau_e)$ (full dot) by a line representing the hybridization function $F_\sigma(\tau_e-\tau_s)$. The black lines correspond to a particle number 1, empty spaces to particle number 0, so the overlaps between the lines for up- and down-spins yield the potential energy. In both approaches, the diagrams corresponding to different connecting ${\cal G}_0$ or $F$ lines are summed up into a determinant and these determinants are sampled by a Monte Carlo procedure.}
\label{diagrams.fig}
\end{figure}

Recently, Rubtsov and co-workers proposed an entirely different approach for solving quantum impurity models \cite{Rubtsov05}. Their continuous-time method is a diagrammatic QMC algorithm which can be regarded as an extension of ideas originally introduced in Ref.~\cite{Prokofev96} to fermionic systems. The algorithm is based on a diagrammatic expansion of the partition function in the interaction term and a stochastic sampling of the resulting diagrams (see Fig.~\ref{diagrams.fig}). More specifically, the action (\ref{S}) is decomposed into a quadratic part
\begin{eqnarray}
S_0 &=& -\int_0^\beta d\tau d\tau' \sum_\sigma c_\sigma(\tau) F_\sigma(\tau-\tau')c_\sigma^\dagger(\tau')\nonumber\\
&&-\mu\int_0^\beta d\tau (n_\uparrow +n_\downarrow)
\label{S_0}
\end{eqnarray}
and an interaction part
\begin{equation}
S_U=U\int_0^\beta d\tau n_\uparrow n_\downarrow.
\label{S_U}
\end{equation}
The weak coupling expansion of $Z=Tr T_\tau e^{-(S_0+S_U)}$ in powers of $U$ then reads
\begin{eqnarray}
Z&=&\sum_k \frac{(-U)^k}{k!}\int d\tau_1 \ldots d\tau_k Tr T_\tau e^{-S_0}\nonumber\\
&&\hspace{20mm}\times n_\uparrow(\tau_1)n_\downarrow(\tau_1)...n_\uparrow(\tau_k)n_\downarrow(\tau_k).
\label{Z_weak}
\end{eqnarray}
The trace over the fermionic degrees of freedom can now be performed analytically. Wick's theorem leads to $2k!$ terms whose combined weight is the determinant of the matrix product $D_{{\cal G}_0,\uparrow}(\tau_1,...,\tau_k)D_{{\cal G}_0,\downarrow}(\tau_1, ..., \tau_k)$. The elements of these $k\times k$ matrices are given by the mean field function defined in Eq.~(\ref{G_0})
\begin{equation}
D_{{\cal G}_0,\sigma}(\tau_1, ..., \tau_k)(i,j) = {\cal G}_{0, \sigma}(\tau_i-\tau_j).
\label{D_G0}
\end{equation}
Finally, the partition function becomes
\begin{equation}
Z=\sum_k \frac{(-U)^k}{k!}\int d\tau_1 \ldots d\tau_k \det \Big[ D_{{\cal G}_0,\uparrow} D_{{\cal G}_0,\downarrow}\Big] 
\label{Z_weak_det}
\end{equation}
and the Monte Carlo sampling proceeds by local updates (random insertions/removals of vertices). At first sight, it appears that the term $(-U)^k$ would lead to a bad sign problem for repulsive interactions. Rubtsov and co-workers found a way to get around this problem by redefining the interaction term $S_U$ with a small positive constant $\alpha$ as 
\begin{eqnarray}
S_U^\alpha &=& \frac{U}{2} \int d\tau \Big[ (n_\uparrow(\tau)+\alpha)(n_\downarrow(\tau)-1-\alpha)\nonumber\\
&& \hspace{11mm} + (n_\uparrow(\tau)-1-\alpha)(n_\downarrow(\tau)+\alpha)\Big]
\label{S_alpha}
\end{eqnarray}
and adjusting the quadratic term $S_0$ in a way to compensate for this change \cite{Rubtsov05}. However, this suppression of the sign problem can lead to a probability distribution $p_\sigma(k)$ in the expansion order which features multiple maxima and the sampling of the different orders then requires some flat-histogram (or similar) method \cite{Rubtsov_private}. 

\subsection{Hybridization expansion}

A complimentary continuous-time algorithm (also illustrated in Fig.~\ref{diagrams.fig}) is obtained by expanding in the hybridization functions $F_\sigma$, while treating the chemical potential and interaction terms exactly. This approach has been worked out in Refs.~\cite{Werner06, Werner06b}. For the hybridization expansion, one decomposes the effective action (\ref{S}) into the (non-local in time) hybridization part
\begin{equation}
S_F = -\int_0^\beta d\tau d\tau' \sum_\sigma c_\sigma(\tau) F_\sigma(\tau-\tau')c_\sigma^\dagger(\tau')
\label{S_F}
\end{equation}
and the local part
\begin{equation}
S_L=-\mu\int_0^\beta d\tau (n_\uparrow +n_\downarrow)+U\int_0^\beta d\tau n_\uparrow n_\downarrow.
\label{S_L}
\end{equation}
Expanding the partition function $Z=Tr T_\tau e^{-(S_F+S_L)}$ in powers of $F_\sigma$ then leads to
\begin{eqnarray}
Z &=& Tr  T_\tau e^{-S_L} \prod_\sigma \sum_{k_\sigma} \frac{1}{k_\sigma!}\int_0^\beta d\tau^\sigma_1... d\tau^\sigma_{k_\sigma}\int_0^\beta d\tilde \tau^\sigma_1 ... d\tilde \tau^{\sigma}_{k_\sigma}\nonumber\\
&& \times \Big[ c_\sigma(\tau_1)F_\sigma(\tau_1-\tilde \tau_1)c^\dagger_\sigma(\tilde \tau_1)\ldots\nonumber\\
&&\hspace{12mm}\ldots c_\sigma(\tau_{k_\sigma})F_\sigma(\tau_{k_\sigma}-\tilde \tau_{k_\sigma})c^\dagger_\sigma(\tilde \tau_{k_\sigma})\Big].
\label{Z_strong}
\end{eqnarray}
The individual terms in this series can have positive or negative sign, but as shown in Ref.~\cite{Werner06}, it is possible to express the combined weight of the $k_\sigma!$ diagrams corresponding to a given collection $\{c^\dagger_\sigma(\tilde \tau_i), c_\sigma(\tau_i)\}_{i=1, \ldots , k_\sigma}$ of creation and annihilation operators as the determinant of a matrix $D_{F, \sigma}$, whose entries are the $F$-functions,
\begin{equation}
D_{F,\sigma}(\tau_1, \ldots, \tau_{k_\sigma}; \tilde{\tau}_1, \ldots, \tilde{\tau}_{k_\sigma})(i,j) = F_\sigma(\tau_i-\tilde\tau_j).
\label{D_F}
\end{equation}
It can be proven in analogy to Ref. \cite{Yoo05} that this diagrammatic formulation does not suffer from a sign problem for models with density-density interactions\footnote{We thank R. Kaul for bringing this to our attention}. The partition function finally becomes
\begin{eqnarray}
Z &=& Tr T_\tau  s_{T_\tau}e^{-S_L}\prod_\sigma \sum_{k_\sigma} \int_{0}^\beta d\tilde\tau^\sigma_1 \int_{\tilde\tau^\sigma_1}^{\beta}d\tau^\sigma_1 \ldots \nonumber\\
&& \hspace{5mm}\ldots \int_{\tilde \tau_{k_\sigma-1}}^{\beta} d\tilde \tau^\sigma_{k_\sigma} \int_{\tilde \tau^\sigma_{k_\sigma}}^{\circ\tilde \tau^\sigma_1}d\tau^\sigma_{k_\sigma}
\det D_{F,\sigma} s_\sigma \nonumber\\
&& \hspace{5mm}\times c_\sigma(\tau^\sigma_{k_\sigma})c^\dagger_\sigma(\tilde \tau^\sigma_{k_\sigma})\ldots c_\sigma(\tau^\sigma_1)c^\dagger_\sigma(\tilde \tau^\sigma_1),
\end{eqnarray}
where $\circ \tau$ denotes an upper integral bound which ``winds around" the circle of length $\beta$.
If the last segment winds around, the sign $s_\sigma$ is $-1$ and otherwise $+1$, whereas $s_{T_\tau}$ compensates for any sign change produced by the time ordering operator. The trace finds an easy and intuitive interpretation in terms of configurations of segments marking the times where a particle of spin $\sigma$ is present \cite{Werner06}. In such a representation, the $\mu$-part of $S_L$ is determined by the total length of the segments while the interaction is given by the total overlap between segments of opposite spin (see Fig.~\ref{diagrams.fig}).

\section{Measuring the Green function}
\subsection{Overview}
\label{measure}

The diagrams obtained from the expansion of the partition function contain vertices or operators which are connected to each other by ``bare" Green functions ${\cal G}_0$ or hybridization functions $F$. In order to measure the Green function, one needs a configuration with two ``unconnected" operators $c^\dagger(\tau)$ and $c(\tau')$. This may either be achieved by inserting such a pair of operators into a given configuration, or removing one of the ${\cal G}_0$- or $F$-functions from an existing diagram. The former approach has been implemented in the weak-coupling algorithm, and the latter in the hybridization expansion algorithm. 

\subsection{Weak Coupling Expansion}
In the case of the weak coupling expansion the diagrams for the Green function $G_{\sigma}(\tau_p- \tau_q)\ = -\langle T_\tau c_\sigma(\tau_p)c_\sigma^\dagger (\tau_q)\rangle$ 
are obtained similarly to the expansion of the partition function (\ref{Z_weak_det}), from the matrix $D_{\mathcal{G}_0,\sigma}^{pq}$ containing an additional row and column: 
\begin{equation}\label{Dpq}
D_{\mathcal{G}_0,\sigma}^{pq}\ =\ \left( \begin{array}{cc} D_{\mathcal{G}_{0,\sigma}} & \mathcal{G}_{0,\sigma}(\tau_i - \tau_q) \\ 
\mathcal{G}_{0,\sigma}(\tau_p - \tau_j) & \mathcal{G}_{0,\sigma}(\tau_p - \tau_q) \end{array} \right).
\end{equation}
The relative weight of the Green function is given by the determinant ratio $\det D^{pq}_{\mathcal{G}_{0,\sigma}}/\det D_{\mathcal{G}_{0,\sigma}},$ which is computed using 
\begin{eqnarray} \label{detD}
\det D_{\mathcal{G}_0}^{pq} &=& \det D_{\mathcal{G}_0} \Big[ \mathcal{G}_0(\tau_p - \tau_q) \nonumber\\
&&- \sum_{ij} \mathcal{G}_0(\tau_p-\tau_i) (D_{\mathcal{G}_0}^{-1})_{ij} \mathcal{G}_0(\tau_j-\tau_q)\Big].\hspace{2mm} 
\end{eqnarray}
Hence the formula for measuring the Green function becomes \cite{Rubtsov05} 
\begin{equation}\label{GFtau}G(\tau_p, \tau_q)\ =\ \mathcal{G}_0(\tau_p-\tau_q) - \Big\langle \sum_{ij} \mathcal{G}_0(\tau_p-\tau_i) M_{ij} \mathcal{G}_0(\tau_j-\tau_q)\Big\rangle, \end{equation}
where $M = D_{\mathcal{G}_0}^{-1}$ and angular brackets denote the Monte Carlo average.  Fourier transforming this formula yields a measurement formula in Matsubara frequencies,
\begin{equation}G(i\omega_n)\ =\ \mathcal{G}_0(i \omega_n) - \Big\langle \beta^{-1}\mathcal{G}_0(i \omega_n)^2 \sum_{ij}M_{ij}e^{i\omega_n(\tau_i-\tau_j)}\Big\rangle.
\label{matsubarameas}
\end{equation}

Both measurements can be performed directly during an update of the partition function, thereby reducing the computational effort for measuring the Green function from $O(N M^2)$ to $O(N M),$ where $N$ is the number of time slices or Matsubara frequencies and $M$ the average matrix size \cite{Rubtsov05}.

The Matsubara Green function is required for the computation of the self-energy and the Hilbert transform, so measuring the Green functions directly in frequency space allows one to avoid the Fourier transformation from imaginary time to Matsubara frequencies. 
In the weak coupling algorithm the measurement in Matsubara frequencies appears as a correction to the (known) bare Green function $\mathcal{G}_0(i\omega_n)$ which is suppressed by a factor of $\frac{1}{\beta i\omega_n}.$ For high frequencies, the errors converge very quickly and it is therefore possible to measure the high frequency behavior in a short time, before focusing on lower Matsubara frequencies for the rest of the simulation. This reduces the computational effort significantly.

\subsection{Hybridization Expansion}

Measurements in the hybridization expansion approach can be performed by removing a hybridization function $F(\tau_i^e-\tau_j^s)$ connecting a pair of creation and annihilation operators. The contribution to the Green function at $\tau=\tau_i^e-\tau_j^s$ is then given by the ratio $(-1)^{i+j} \det D^{ij}_F/ \det D_F$, where $D_F^{ij}$ denotes the the matrix $D_F$ with the row $i$ and column $j$ removed. Since 
\begin{eqnarray}
(-1)^{i+j}\det D^{ij}_F = \det D_F (D_F^{-1})_{ji}
\end{eqnarray}
it follows that the Green function can be measured from the inverse hybridization matrix $M=D_F^{-1}$ as
\begin{equation}
\label{strongmeas} G(\tau)\ =\Big \langle \frac{1}{\beta}\sum_{ij} M_{ji} \Delta(\tau, \tau_i^e-\tau_j^s)\Big \rangle, 
\end{equation}
 where 
 $\Delta(\tau, \tau') = \operatorname{sgn}(\tau') \delta(\tau-\tau'-\theta(-\tau')\beta).$
 
This yields $O(M^2)$ (correlated) estimates for the Green function in one step. For $\tau$ near 0 or $\beta$, the Green function converges rapidly, whereas the variance of these measurement values is relatively large for $\tau \approx \beta/2$. The number of imaginary time slices does not influence the performance of the algorithm, and thus can be chosen arbitrarily large. It is therefore possible to adjust the resolution after the simulation according to the slope of the Green function and the noise in the measured data.   

Similar to the weak coupling case, formula (\ref{strongmeas}) could be Fourier transformed and $G(i\omega_n)$ measured directly. However, this approach is not advantageous, because the computation of the exponentials is very expensive compared to the rest of the simulation, and the binned values can easily be computed on a fine grid. Since the measured values are in any case not a small correction to a known function, the hybridization expansion algorithm has more difficulties obtaining accurate results for the high-frequency behavior. 

As noted in Ref.~\cite{Werner06}, the semi-circular density of states is a special case, where the self-consistency loop can be performed without any Fourier-transformation. The hybridization expansion results presented in the following section, however, were always obtained in the ``conventional" way, by Fourier-transforming the Green function $G(\tau)$ and extracting the self-energy.

\begin{figure}[ht]
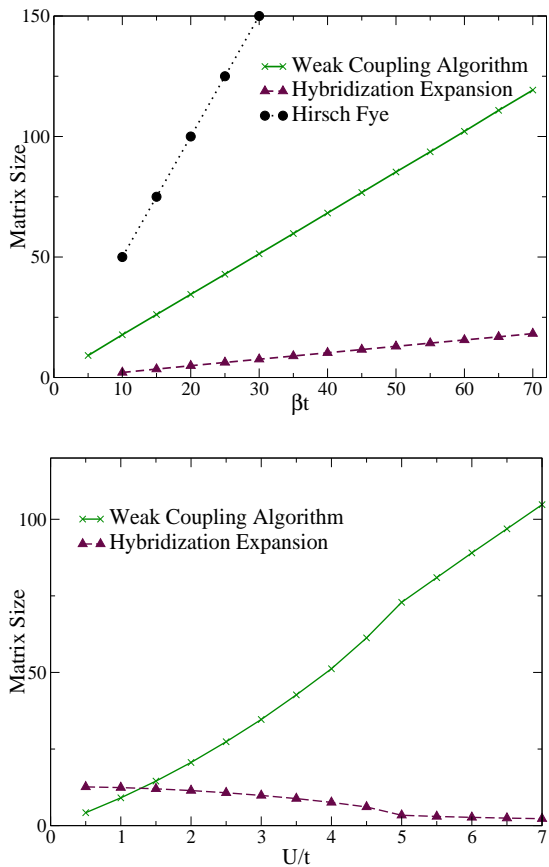

\begin{center}
\includegraphics[width=.4\textwidth,height=0.23\textheight]{MatrixSizes.eps} \\ 
%
\vspace{5mm}
\includegraphics[width=.4\textwidth,height=0.23\textheight]{MatrixU.eps} \\ 
\end{center}
\caption[MatrixU]{Scaling of the matrix size with inverse temperature and interaction strength. Upper panel: temperature dependence for $U/t=4$. In the case of Hirsch-Fye, the resolution $N=\beta U$ has been chosen as a compromise between reasonable accuracy and acceptable speed, while the average matrix size is plotted for the continuous-time solvers. Lower panel: dependence on $U/t$ for fixed $\beta t=30$. The solutions for $U \leq 4.5$ are metallic, while those for $U \geq 5.0$ are insulating. The much smaller matrix size in the relevant region of strong interactions is the reason for the higher efficiency of the hybridization expansion method.}
\label{Matrix.fig}
\end{figure} 

\subsection{Susceptibilities}
In the weak coupling algorithm four-point functions 
$G^{(2)}_{\sigma \sigma'}(\tau_1, \tau_2, \tau_3, \tau_4) = \langle T_\tau c_{\sigma}^\dagger(\tau_1) c_{\sigma}(\tau_2) c_{\sigma'}^\dagger(\tau_3) c_{\sigma'}(\tau_4)\rangle$
can be measured by inserting operator pairs in analogy to the Green function (\ref{Dpq}). 
For same spin $\sigma=\sigma'$ we then have to compute the determinant ratio of a matrix with two added rows and two added columns.
For opposite spins one finds the product of two determinant ratios of the form (\ref{detD}).

In the hybridization expansion approach the four point functions are obtained by removing two hybridization lines, which leads to the measurement of 
$\sum_{ijkl}(M_{\sigma})_{ij}(M_{\sigma'})_{kl}$ for 
$\sigma \neq \sigma'$ and $\sum_{ijkl}[(M_\sigma)_{ij}(M_\sigma)_{kl}-(M_\sigma)_{il}(M_\sigma)_{kj}]$ for equal spins. Spin and charge susceptibilities, 
or more generally the density-density correlators can be obtained independently and accurately (and with negligible computational effort) from the segment configurations. These segments, shown in the lower panel of Fig. \ref{diagrams.fig}, represent the occupation of the orbital.

\section{Results}

\subsection{Matrix size}

For all three algorithms, the computational effort scales as the cube of the matrix size, which for the Hirsch-Fye solver is determined by the time discretization $\Delta\tau=\beta/N$ and in the case of the continuous-time solvers is determined by the perturbation order $k$, which is peaked roughly at the mean value determined by the probability distribution $p(k)$. In Fig.~\ref{Matrix.fig}, we plot these matrix sizes as a function of inverse temperature $\beta$ for fixed $U/t=4$ and as a function of $U/t$ for fixed $\beta t=30$. All our simulation results are for a semi-circular density of states with band-width $4t$.

It is obvious from the upper panel of Fig.~\ref{Matrix.fig} that the matrix size in all three algorithms scales linearly with $\beta$. The Hirsch-Fye data are for $N=\beta U$, which is apparently a common choice, although Figs.~\ref{e_kin.fig} and \ref{sigma_lowest_freq.fig} show that it leads to considerable systematic errors. Thus, the grid size should in fact be chosen much larger ($N\gtrsim5\beta U$).

While the matrix size in the weak coupling approach is approximately proportional to $U/t$, as in Hirsch-Fye, the $U$-dependence of the hybridization expansion algorithm is very different: a decrease in average matrix size with increasing $U/t$ leads to much smaller matrices in the physically interesting region $4\lesssim U/t \lesssim 6$, where the Mott transition occurs. The results in Fig.~\ref{Matrix.fig} and the cubic dependence of the computational effort on matrix size essentially explain why the continuous-time solvers are much more powerful than Hirsch-Fye and why the hybridization expansion is best suited to study strongly correlated systems. 

There is of course a prefactor to the cubic scaling, which depends on the computational overhead of the different algorithms and on the details of the implementation. Bl\"{u}mer \cite{Blumer} has demonstrated substantial optimizations of the Hirsch-Fye code and has in particular shown that extrapolating results at non-zero time step $\Delta\tau$ to the $\Delta\tau=0$ limit considerably improves the accuracy. Of the continuous time codes investigated here, only the weak coupling results have been optimized. We estimate that similar modifications in the code for the hybridization expansion algorithm would provide a speed-up of at least a factor of 10. However, the results presented here indicate large enough difference between the methods that the effects of optimization can be ignored. 

\subsection{Accuracy for constant CPU time}

The three quantum Monte Carlo algorithms considered in this study work in very different ways. Not only are the configuration spaces and hence the update procedures entirely different, but also the measurements of the Green functions and other observables. 

In order to study the performance of the different impurity solvers, we therefore decided to measure the accuracy to which physical quantities can be determined for fixed CPU time (in this study 7h on a single Opteron 244 per iteration). This is the question which is relevant to people interested in implementing either of the methods and avoids the tricky (if not impossible) task of separating the different factors which contribute to the uncertainty in the measured results. Because the variance of the observables measured in successive iterations of the self-consistency loop turned out to be considerably larger than the statistical error bars in each step, we determined the mean values and error bars using 20 DMFT iterations starting from a converged solution.  

The Hirsch-Fye solver suffers in addition to these statistical errors from systematic errors due to time discretization. These systematic errors are typically quite substantial and much larger than the statistical errors. In order to extract meaningful results from Hirsch-Fye simulations it is essential to do a careful (and time-consuming) $\Delta\tau\rightarrow 0$ analysis \cite{Blumer}. The continuous-time methods are obviously free from such systematic errors if a sufficient number of time- or frequency points is used in the measurement of the Green function.   


\subsubsection{Kinetic and potential energy}

\begin{figure}[ht]
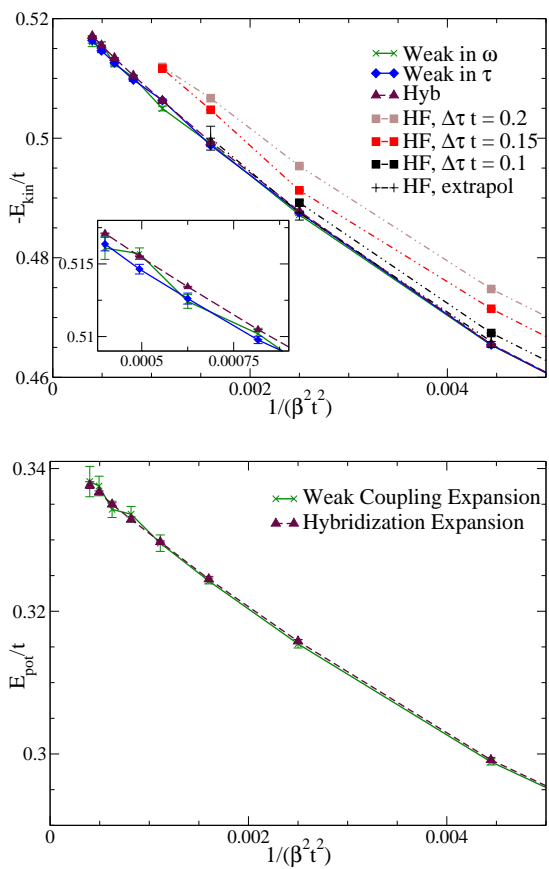

\begin{center}
\includegraphics[width=.4\textwidth,height=0.23\textheight]{e_kin.eps} \\
\vspace{5mm} 
\includegraphics[width=.4\textwidth,height=0.23\textheight]{e_pot.eps} \\ 
\end{center}
\caption[]{Upper panel: kinetic energy $E_{kin}=2 t^2 \int_0^\beta d\tau G(\tau)G(-\tau)$ obtained using the three QMC impurity solvers for $U/t = 4.0$ and $\beta t = 10, 15, \ldots, 50$. The Hirsch-Fye simulations for $\Delta \tau =1/U$ (as in Fig. \ref{Matrix.fig}) yield systematically higher energies. The inset shows results obtained with the continuous-time solvers for $\beta t=35, 40, 45$ and 50. Lower panel: potential energy $U\langle n_\uparrow n_\downarrow \rangle$ for the same interaction strength. 
} 
\label{e_kin.fig}
\end{figure} 
The kinetic energy, 
\begin{equation}
E_\text{kin}=2 t^2 \int_0^\beta d\tau G(\tau)G(-\tau),
\label{e_kin}
\end{equation}
shown in  Fig.~\ref{e_kin.fig}, was obtained from the imaginary time Green function by numerical integration. To this end we Fourier transformed the imaginary time Green function and summed the 
frequency components including the analytically known tails. This turns out to be more accurate than the direct evaluation of equation (\ref{e_kin}) by trapezoidal or Simpson rule. It is also more accurate
than the procedure proposed in Ref. \cite{Haule07} for the temperature and interaction range studied. 

We computed results for fixed $U/t=4$ and temperatures  $\beta t=10, 15, \ldots, 50$. In this parameter range the solution is metallic and we expect $E_\text{kin}/t\propto (T/t)^2$ at low temperature. The dominant contribution to $E_\text{kin}$ comes from imaginary time points close to $\tau=0, \beta.$ The accuracy of the kinetic energy therefore illustrates how well the steep initial drop of $G(\tau)$ can be resolved.

The results from the continuous-time solvers agree within error bars, but due to the larger matrix size, the weak coupling algorithm can perform fewer updates for fixed CPU time and therefore the error bars are substantially larger (see inset of Fig.~\ref{e_kin.fig}). 

The Hirsch-Fye results are strongly dependent on the number of time slices used. Because of the cubic scaling of the computational effort with the number of time slices, at most a few hundred time points can be taken into account. This number is not sufficient to resolve the steep drop of the Green function at low temperature, and therefore the kinetic energy converges to values which are systematically too high.  Extrapolation (e.g. ref. \cite{Georges96}, \cite{Blumer}) can be used to obtain values for $\Delta\tau = 0$ and reduce these errors. However, various simulations at different $\Delta\tau$ have to be performed in order to obtain an accurate estimate.
For the kinetic energy we performed this extrapolation for $\beta t\ =\ 15, 20, 25$. The error for $\beta\ t =\ 20$ at $\Delta \tau\ =\ 0$ after extrapolation is 10 times larger than the one we could obtain for the weak coupling algorithm, which is again around ten times larger than the one for the hybridization algorithm.

We emphasize that for this particular case all three methods are sufficiently accurate that physically meaningful conclusions can be drawn; the differences, however, have clear implications for the extension of the method to more demanding regimes.

In the lower panel of Fig.~\ref{e_kin.fig} we show the potential energy $U\langle n_\uparrow n_\downarrow \rangle$ for $U/t=4$, computed with the two continuous-time methods. In the hybridization expansion algorithm, the double occupancy can be measured from the overlap of the up- and down-segments. In the weak-coupling case, we used the relation $U/2 \langle (n_\uparrow+\alpha)(n_\downarrow-1-\alpha) + (n_\uparrow-1-\alpha)(n_\downarrow+\alpha)\rangle = \langle k \rangle/\beta$ (where $\langle k\rangle $ is the average perturbation order), and an extrapolation to $\alpha\rightarrow 0$. Both results agree within error bars and the hybridization expansion approach again yields the more accurate results.

\begin{figure}[t]
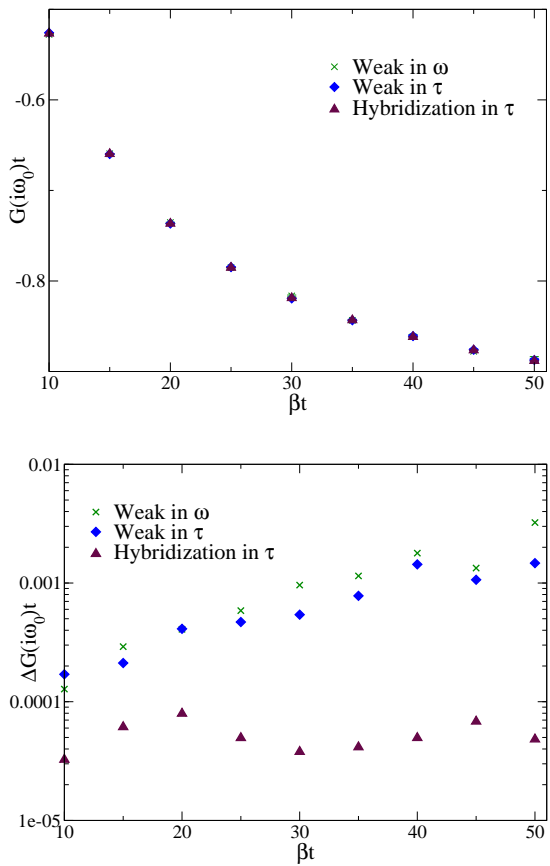

\begin{center}
\includegraphics[width=.4\textwidth,height=0.23\textheight]{lfreq.eps} \\ 
\vspace{5mm}
\includegraphics[width=.4\textwidth,height=0.23\textheight]{lfreq_err.eps} \\ 
\end{center}
\caption[]{ Lowest Matsubara frequency value for $G$ for $U/t=4.0,$ using measurements in both imaginary time and frequency space in the weak coupling case.
The upper panel shows the Green function and the lower panel the relative error on the measurement. Unlike in the Hirsch Fye algorithm there are essentially no systematic errors in the continuous time algorithms. In the case of the hybridization expansion algorithm, results for measurements in $\tau$ and $\omega$ are plotted. Both measurements yield a similar accuracy at low frequency. The hybridization expansion algorithm gives very accurate results and the error bars show no dependence on $\beta$. This indicates that in the measured temperature range, two competing effects essentially cancel: the efficiency of the matrix updates which decreases at lower temperatures and the efficiency of the measurement procedure (\ref{strongmeas}), which yields better results for larger matrix sizes.
}
\label{lfreq.fig}
\end{figure} 

\subsubsection{Green function and self energy}

The high precision of the hybridization expansion results for the kinetic energy indicate that this algorithm can accurately determine the shape of the Green function near $\tau=0$ and $\beta$. We now turn to the lowest Matsubara frequency component of the Green function, which is determined by the overall shape. We plot in Fig.~\ref{lfreq.fig} $G(i\omega_0)$ for different values of $\beta$. The upper panel shows the results obtained for the different continuous-time solvers and measurement procedures. They all agree within error bars. In the lower panel we plot the values of the error-bars. In the case of the weak-coupling expansion, both the measurement in $\tau$ and the measurement in $\omega$ produce about the same accuracy, which deteriorates as the temperature is lowered, due to the increasing matrix size. The error-bars from the hybridization expansion solver are much smaller and in the measured temperature range remain about constant. Because the matrices at these values of $U$ and $\beta$ are very small, and the number of measurement points in Eq.~(\ref{strongmeas}) depends on the matrix size, the increase in computer time for updating larger matrices is compensated by a more efficient measurement.  

\begin{figure}[t]
\begin{center}
\includegraphics[width=.4\textwidth,height=0.23\textheight]{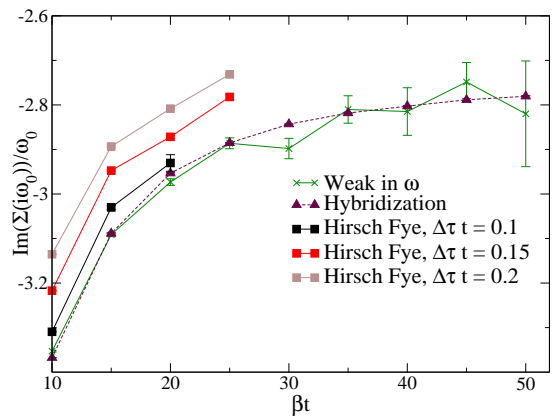} \\ 
\end{center}
\caption[]{Self-energy $\Im m\Sigma(i\omega_0)/\omega_0$ as a function of $\beta$ for $U/t=4.0$. The Hirsch-Fye results exhibit large discretization errors, while the continuous-time methods agree within error bars. The hybridization expansion method is particularly suitable for measuring quantities which depend on low-frequency components, such as the quasi-particle weight. }
\label{sigma_lowest_freq.fig}
\end{figure} 


For the self-energy,
\begin{equation} \Sigma(i\omega_n)\ =\ \mathcal{G}_0(i\omega_n)^{-1}-G(i\omega_n)^{-1},\end{equation}
the Matsubara Green functions have to be inverted and subtracted. This procedure amplifies the errors of the self-energy 
especially in the tail region where $\mathcal{G}_0(i\omega_n)$ and $G(i\omega_n)$ have similar values. Fig.~\ref{sigma_lowest_freq.fig} shows $\Im m \Sigma(i\omega_0)/\omega_0$ for $U/t=4$ and several values of $\beta$. This quantity is related to the quasi-particle weight $Z\approx 1/(1-\Im m \Sigma(i\omega_0)/\omega_0)$. Again, the Hirsch-Fye results show large systematic errors due to the time discretization and cannot be carried to low temperatures. The results from the continuous-time solvers agree within error-bars, but the size of the error bars is very different. The hybridization expansion approach yields very accurate results for low Matsubara frequencies in general.

\begin{figure}[t]
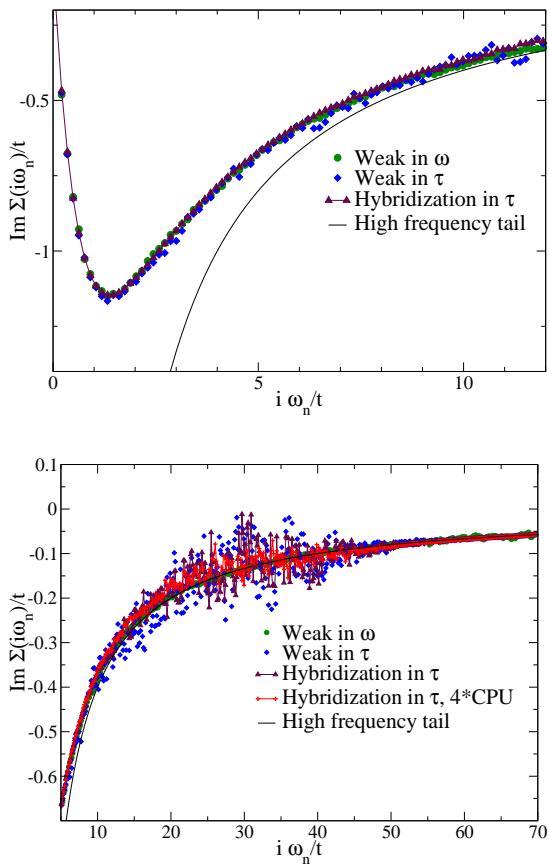

\begin{center}
\includegraphics[width=.4\textwidth,height=0.23\textheight]{selfenergy_tau_omega2.eps} \\ 
\vspace{5mm}
\includegraphics[width=.4\textwidth,height=0.23\textheight]{selfenergy_tau_omega.eps} \\ 
\end{center}
\caption[]{ 
Upper panel: low frequency region of the self-energy $\Sigma(i\omega)$ for $U/t=4.0, \beta t=45.$ Noise in the higher frequencies is clearly visible for the values measured in $\tau$, while the values measured in $\omega$ in the weak coupling algorithm converge smoothly to the high frequency tail.
Lower panel: high frequency region of the self-energy $\Sigma(i\omega)$ for $U/t=4.0, \beta t=45.$ Noise in the higher frequencies is clearly visible for the values measured in $\tau$, while the values measured in $\omega$ in the weak coupling algorithm converge smoothly to the high frequency tail, $\lim_{\omega \rightarrow 0} \Sigma(i\omega_n) = U^2(1-n)n/(i\omega_n)$.}
\label{selfenergy_tau_omega.fig}
\end{figure} 


\begin{figure}[ht]
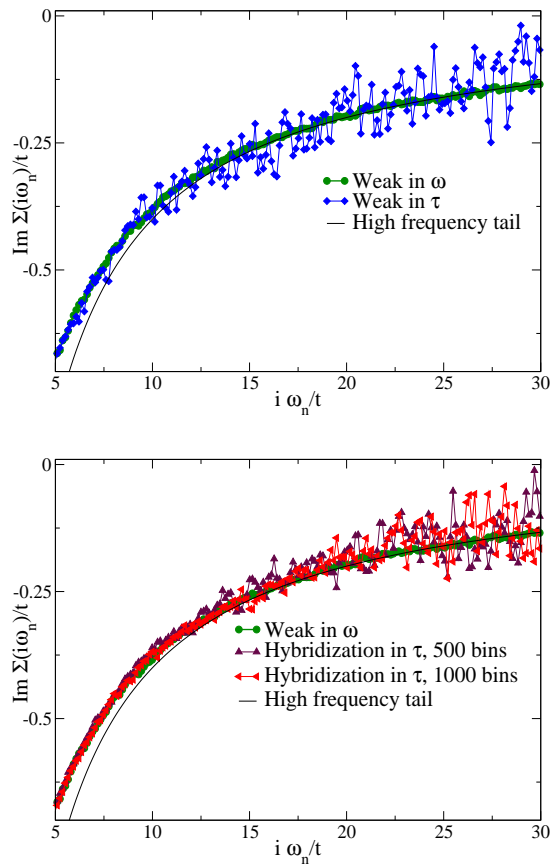

\begin{center}
\includegraphics[width=.4\textwidth,height=0.23\textheight]{selfenergy_tau_omega_rubtsov.eps} \\ 
\vspace{5mm}
\includegraphics[width=.4\textwidth,height=0.23\textheight]{selfenergy_tau_omega_werner.eps} \\ 
\end{center}
\caption[]{Intermediate frequency region of the self-energy $\Sigma(i\omega)$ for $U/t=4.0, \beta t=45$. Noise in the higher frequencies is clearly visible for the values measured in $\tau$, while the values measured in $\omega$ in the weak coupling algorithm converge smoothly to the high frequency tail.}
\label{selfenergy_tau_omega_rubtsovwerner.fig}
\end{figure}

The advantage of measuring in Matsubara frequencies as opposed to imaginary time in the weak coupling algorithm becomes apparent for large $\omega_n$. Only the difference of $G$ to the bare Green function $\mathcal{G}_0$ has to be measured in this algorithm. These differences 
decrease with $1/\omega_n$ for increasing $\omega_n$ and the estimate from Eq.~(\ref{matsubarameas}) is extremely accurate at high frequencies, so that the tail of the self energy can be computed accurately. The measurements in imaginary time however have to be binned and Fourier transformed. While the high frequency tail can be enforced using correct boundary conditions for the cubic splines, there is a region of frequencies which starts much below the Nyquist frequency, where this  introduces considerable errors (Fig.~\ref{selfenergy_tau_omega.fig}). For $10 \lesssim \omega_n/t \lesssim 40$ and $500$ imaginary time slices the values of $\Sigma(i\omega_n)$ show large errors before converging to the high-frequency tail enforced by the Fourier transformation procedure. The upper panel of Fig.~\ref{selfenergy_tau_omega_rubtsovwerner.fig} shows the difference between the two measurement approaches more clearly. 

The hybridization expansion algorithm starts from the atomic limit and thus does not get the high-frequency tail automatically right. Both a measurement in $\tau$ and $\omega$ leads to relatively large errors at high frequencies. This noise again sets in at frequencies much below the Nyquist frequency, as illustrated by the results for 500 and 1000 bins in the lower panel of Fig.~\ref{selfenergy_tau_omega_rubtsovwerner.fig}. This noise is the consequence of the statistical errors in the Green function and can hence be reduced by running the simulation for a longer time (see Fig.~\ref{selfenergy_tau_omega.fig}). However, Fig.~\ref{selfenergy_tau_omega_rubtsovwerner.fig} also shows that even for the shorter runs, the data remain accurate up to sufficiently large $\omega_n$ that a smooth patching onto the analytically known high-frequency tail appears feasible. Furthermore, since the hybridization expansion results in this section have all been obtained without any patching or smoothing and nicely agree with those from the weak-coupling solver, it seems that this uncertainty in the high-frequency tail is not a serious issue.

\begin{figure}[t]
\begin{center}
\includegraphics[width=.4\textwidth,height=0.23\textheight]{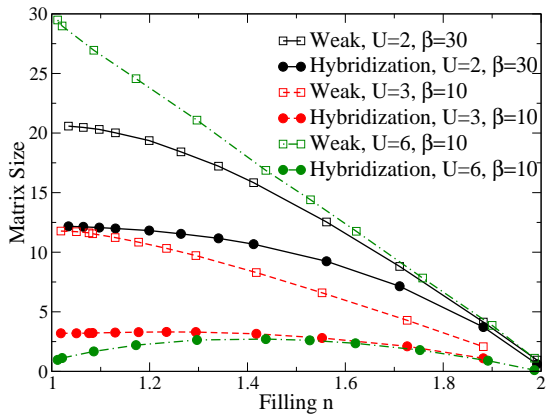} \\ 
\end{center}
\caption[]{ Matrix sizes away from half filling: the matrix size decreases for the weak coupling algorithm, while the one for the hybridization expansion algorithm increases as one dopes a Mott-insulating state.}
\label{matrix_size_U2_beta30.fig}
\end{figure}

\subsection{Away from half filling}

We have tested both continuous time algorithms away from half filling, in a region where the half-filled model at zero temperature has a gap ($U/t = 6, \beta t=10$) and in a region without gap ($U/t=3, \beta t=10$, $U/t=2, \beta t=20$). A comparison of the Green functions and self-energies has shown that both algorithms produce the same result within numerical precision and are much faster than Hirsch-Fye. Both continuous time algorithms have no sign problem away from half filling (\cite{Yoo05}, \cite{Rubtsov05}) and again the time needed to obtain a given accuracy is mostly determined by the size of the matrix. In the case of the weak coupling algorithm it decreases continuously away from half filling, while in the case of the hybridization expansion the perturbation order first increases with doping if the half-filled model has a gap, and then decreases (see Fig.~\ref{matrix_size_U2_beta30.fig}). For all regions of parameter space tested, the hybridization expansion approach yields the smaller matrix sizes and is therefore substantially faster. The matrix sizes become comparable only in the limit of filled or empty bands.

For the hybridization expansion algorithm, we have also computed the matrix size for $U/t=6$ and much lower temperatures $\beta t=100$, 200 and 400. These results showed that the perturbation order for a given filling remains proportional to $\beta$, so that the shape of the curve remains the same as shown for $\beta t=10$ in Fig.~\ref{matrix_size_U2_beta30.fig}. In particular this means that the formation of the ``Kondo resonance" (which contains the physics of coherent low energy quasi-particles) in the slightly doped system at low temperatures does not lead to any dramatic change in the perturbation order.

\section{Conclusions}  

We compared the performance of three different impurity solvers: the Hirsch-Fye auxiliary field approach which for the last 15 years has been widely used in DMFT calculations, and two recently developed continuous-time algorithms, which are based on a stochastic sampling of an expansion of the partition function in the interaction and hybridization, respectively. Both continuous-time methods were found to be much more efficient than Hirsch-Fye for all relevant values of temperature, interaction strength and filling. Because the time-discretization in Hirsch-Fye simulations furthermore introduces a systematic error which can be substantial and is difficult to estimate without a careful analysis involving several runs for different $\Delta\tau$, it makes sense (in most cases) to replace the method by one of the continuous-time solvers. 

The hybridization expansion leads to much smaller matrix sizes at intermediate and strong couplings, than the weak-coupling expansion. Hence it allows access to much lower temperatures. Our analysis has shown that the hybridization expansion is particularly powerful at calculating low-frequency components of the Green function and quantities which are sensitive to these components, such as the quasi-particle weight. The method has more difficulties capturing the high-frequency behavior correctly. In the weak-coupling approach, the algorithm perturbs around the non-interacting solution, which has the correct high-frequency tail and if the Green function is measured directly in frequency space, the intermediate and high Matsubara frequencies can be determined very accurately. Its power has also been demonstrated with recent applications to the Kondo lattice, multi orbital models \cite{Werner06b}, and CDMFT \cite{Haule07}. 
 
We do not expect the noise in the high frequency components to be a serious problem, because the noise appears at high enough frequencies that a smooth patching onto the analytically known tail seems feasible. However, the application of the hybridization expansion algorithm to a model with arbitrary density of states will be an important test. Efforts to use it in the simulation of vanadium oxide are under way and results will be presented elsewhere.  

\acknowledgments

PW and AJM acknowledge support from NSF DMR 0431350. The calculations have been performed on the Hreidar Beowulf cluster at ETH Z\"urich, using the ALPS library \cite{ALPS}. We thank F.~F.~Assaad, N.~Bl\"{u}mer and A.~N.~Rubtsov for helpful discussions.

\end{document}